# MODELING OF MAGNETOELECTRIC EFFECTS IN FERROMAGNETIC/PIEZOELECTRIC BULK COMPOSITES

V.M. PETROV, M.I. BICHURIN V. M. LALETIN*,
N.N.PADDUBNAYA * AND G. SRINIVASAN**

*Novgorod State University, B. S. Peterburgskaya St. 41, 173003 Veliky Novgorod, Russia
*Institute of Technical Acoustics, 210717 Vitebsk, Belarus,
**Physics Department, Oakland University, Rochester, MI 48309 USA*

### ABSTRACT

We discuss a model that considers the bulk composite as a homogeneous medium with piezoelectric and magnetostrictive subsystems. We solve combined elastostatic, electrostatic and magnetostatic equations to obtain effective composite parameters (piezoelectric modules, magnetostriction factors, compliances, magnetoelectric coefficients) for 3-0 and 0-3 connectivities. Expressions for longitudinal and transverse low-frequency magnetoelectric voltage coefficients have been obtained for unclamped and clamped samples. Volume fractions for peak low-frequency effective magnetoelectric voltage coefficient are found to be dependent on specific connectivity. Clamping leads to significant variation in magnetoelectric voltage coefficients. The calculated magnetoelectric coefficients are compared with data.

## 1. INTRODUCTION

A bulk composite consisting of ferrite-ferroelectric phases is expected to be magnetoelectric (ME) since the ME coefficient $\alpha_E = \delta E/\delta H$ is the product of the magnetostrictive deformation $\delta z/\delta H$ and the piezoelectric field generation $\delta E/\delta z$. Bulk composites are desirable over layered samples due to superior mechanical strength. One could also easily control physical, magnetic, electrical and ME parameters with proper choice for the two phases and their volume fraction.

Harshe, et al., performed calculations of ME coefficients in bulk samples for 3-0 and 0-3 connectivities.[1] They assumed that the composite magnetic field is same as the internal field in the ferrite. ME coefficients, therefore, were found in terms of the ratio of external electric field to magnetic field in ferrite component i.e. $E_3/{}^mH_3$. However, the average magnetic field in composite sample differs from magnetic field in ferrite. Getman developed the theory of ME effect in bulk composites for 3-1 connectivity that is extremely difficult to realize.[2] The purpose of the present work is the development of a general theory describing ME interactions in composites for 3-0 and 0-3 connectivities. The present mode allows the determination of effective composite parameters including ME susceptibility and ME coefficients.

## 2. GENERAL APPROACH

Let us consider, for example, a composite with 3-0 connectivity. Dimensions of sample are supposed to be small compared with wave-lengths of ac fields involved in the measurements. The sample shape is assumed to be cubes as shown in Fig. 1. One, therefore, needs to obviously analyze only one of the units to describe the whole sample.

A two stage averaging procedure is used for obtaining the effective composite parameters.[3] In the first stage, the composite is considered as a structure consisting of piezoelectric and magnetostrictive phases.

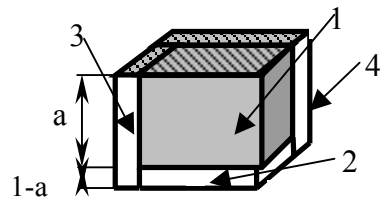

*Figure 1. Cube model of composite with connectivity 3-0:*
*1 is magnetostictive phase;*
*2, 3 and 4 are piezoelectric phase.*

For the polarized piezoelectric phase with the symmetry ∞m and magnetostrictive phase with the cubic symmetry, the following equations can be written for the strain and electric and magnetic displacement:

$$^pS_i = {}^ps_{ij}\,{}^pT_j + {}^pd_{ki}\,{}^pE_k;$$
$$^pD_k = {}^pd_{ki}\,{}^pT_i + {}^p\varepsilon_{kn}\,{}^pE_n;$$
$$^mS_i = {}^ms_{ij}\,{}^mT_j + {}^mq_{ki}\,{}^mH_k;$$
$$^mB_k = {}^mq_{ki}\,{}^mT_i + {}^m\mu_{kn}\,{}^mH_n;$$

where $^pS_i$, $^pT_j$, $^mS_i$ and $^mT_j$ are strain and stress tensor components of the piezoelectric and magnetostrictive phases, $^pE_k$ and $^pD_k$ are the vector components of electric field and electric displacement, $^mH_n$ and $^mB_k$ are the vector components of magnetic field and magnetic induction, $^ps_{ij}$, $^ms_{ij}$, $^pd_{ki}$, and $^mq_{ki}$ are compliance of the piezoelectric and magnetostrictive phases, piezoelectric and piezomagnetic coefficients, $^p\varepsilon_{kn}$ is the permittivity matrix and $^m\mu_{kn}$ is the permeability matrix.

In Eq. (1) the gravitational force is small compared to internal forces. In the second stage, the bilayer is considered as homogeneous[3] and the behavior is described by:

$$S_i = s_{ij}T_j + d_{ki}E_k + q_{ki}H_k;$$
$$D_k = d_{ki}T_i + \varepsilon_{kn}E_n + \alpha_{kn}H_n;$$
$$B_k = q_{ki}T_i + \alpha_{kn}E_n + \mu_{kn}H_n,$$

where $S_i$ and $T_j$ are the average strain and stress tensor components, $E_k$, $D_k$, $H_k$, and $B_k$ are the average vector components of electric field, electric displacement, magnetic field and magnetic induction, $s_{ij}$, $d_{ki}$, and $q_{ki}$ are effective compliance, piezoelectric and piezomagnetic coefficients, and $\varepsilon_{kn}$, $\mu_{kn}$ and $\alpha_{kn}$ are effective permittivity, permeability and ME coefficient. Effective parameters of the composite are obtained by solving Eq. (2), taking into account solutions of Eq. (1) and boundary conditions.

In the case of longitudinal ME effect the composite is poled with an electric field $E$ along direction *3*. The bias field $H$ and the ac field $\delta H$ are along the same direction as $E$ and the resulting induced electric field $\delta E$ is estimated across the sample thickness. Then we find an expression for $\alpha'_{E,L} = \alpha'_{E,33} = \delta E_3/\delta H_3$. Equations (1) and (2) are then solved for the following boundary conditions:

$^1S_1 = {}^2S_1,$
$^1S_2 - {}^2S_2 = 0,$
$^2S_1 - {}^3S_1 = 0,$
$^3S_3 - {}^4S_3 = 0,$
$a^1S_3 + (1-a)^2S_3 - {}^3S_3 = 0,$
$a^2E_3 + (1-a)^3S_2 - {}^4S_2 = 0,$
$^1A_1(T) + {}^2A_1{}^2T_1 + {}^3A_1{}^3T_1 = 0,$
$^4T_1 = 0,$
$^3A_2{}^3T_2 + {}^4A_2{}^4T_2 = 0,$

$^1A_2{}^1T_2 + {}^2A_2{}^2T_2 = {}^3A_2{}^1T_2,$
$^1T_3 - {}^2T_3 = 0,$
$^2A_3{}^2T_3 + {}^3A_3{}^3T_3 + {}^4A_3{}^4T_3 = 0,$
$a^1E_3 + (1-a)\,{}^2E_3 - {}^3E_3 = 0,$
$^3E_3 - {}^4E_3 = 0,$
$^1D_3 - {}^2D_3 = 0,$
$^2A_3{}^2D_3 + {}^3A_3{}^3D_3 + {}^4A_3{}^4D_3 = D_3,$
$a^1H_3 + (1-a)\,{}^2H_3 - {}^3H_3 = 0,$
$^3H_3 - {}^4H_3 = 0,$
$^1B_3 - {}^2B_3 = 0,$
$^2A_3{}^2B_3 + {}^3A_3{}^3B_3 + {}^4A_3{}^4B_3 = B_3,$

where $^iS_j$, $^iT_j$, $^iE_j$, $^iD_j$, $^iH_j$, $^iB_j$ denote strain, stress, electric field and displacement, magnetic field and induction components, respectively, and $^iA_j$ is cross-section area of the $i^{th}$ unit perpendicular to direction $j$.

(2)

By combining the solutions of Eqs. (1) and (2) and taking into account Eq. (3), one finds the effective composite parameters for unclamped samples with 3-0 connectivity in the longitudinal case. These parameters are found also for clamped sample using expression $S_3 = -s_c s_{33}T_3$. Here $s_c$ is relative compliance of clamp system. The transverse case is considered by the same procedure. Effective parameters for 0-3 connectivity are obtained similarly; the 0-3 connectivity differs from 3-0 connectivity in that the ferroelectric and ferrite phases are interchanged. Analytical expressions for effective composite parameters are too tedious; it is more practical to solve these equations numerically by computer.

## 3. COMPARISON WITH DATA AND DISCUSSION

Bulk composites of lead zirconate titanate (PZT) with NiFe₂O₄ (NFO) is investigated. For computations the parameters of the two phases used for calculations are as follows:

*PZT*:
$^p s_{11}$ = 15.3·10$^{-12}$ m$^2$/ N; $^p s_{12}$ = -5·10$^{-12}$ m$^2$/ N; $^p s_{13}$ = -7.22·10$^{-12}$ m$^2$/ N; $^p s_{33}$ = 17.3·10$^{-12}$ m$^2$/ N; $^p d_{31}$ = -175·10$^{-12}$ m/V; $^p d_{33}$ =400·10$^{-12}$ m/V, $^p \varepsilon_{33}/\varepsilon_0$=1750.

*NiFe$_2$O$_4$*:
$^m s_{11}$ = 6.5·10$^{-12}$ m$^2$/ N; $^m s_{12}$ = -2.4·10$^{-12}$ m$^2$/ N; $^m q_{31}$ = 125·10$^{-12}$ m/A;
$^m q_{33}$ = --680·10$^{-12}$ m/A, $^m \mu_{33}/\mu_0$=3, $^m \varepsilon_{33}/\varepsilon_0$=10.

Calculations show that the peak longitudinal ME voltage coefficient for 3-0 connectivity reaches 4000 mV/cm Oe and is three times higher than the transverse coefficient. In case of 0-3 connectivity the peak longitudinal ME voltage coefficient equals 900 mV/cm Oe. Clamping leads to significant decrease of ME voltage coefficients. In a real composite the internal units are clamped by neighboring ones. Therefore the ME voltage coefficient calculated by Harshe et al., for unclamped sample significantly exceeded the measured value. Dependence of transverse ME voltage coefficients on volume fraction of ferroelectric phase *v* are shown in Fig. 2 for clamped sample ($s_c = 0.3$). Data are from Ref. 4.

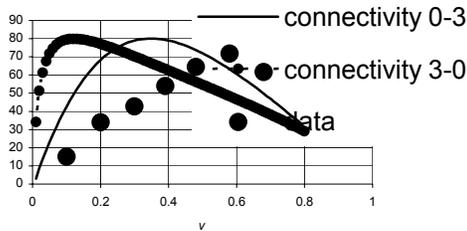

*Figure 2. Dependence of ME voltage transverse coefficient on volume fraction of ferroelectric phase v for bulk composite of NFO-PZT*

Figure 2 shows that peak ME voltage coefficient corresponds to a volume fraction of ferroelectric phase of 0.11 for 3-0 connectivity and is 0.36 for 0-3 connectivity while measured peak ME voltage coefficient is observed at volume fraction 0.6. On that evidence one can conclude that the 0-3 connectivity is the most appropriate for describing the ME effect in the composite. The discrepancy between theory and data is probably due to variation of actual material parameters from the assumed values.

## 4. CONCLUSION

Solutions for longitudinal and transverse low-frequency ME voltage coefficients have been obtained for unclamped and clamped samples. It is shown that for 3-0 connectivity the longitudinal ME voltage coefficient is three times higher than the transverse coefficient for bulk composite of lead zirconate titanate with NiFe$_2$O$_4$. Volume fractions corresponding to peak ME voltage coefficients depend on connectivity. It is shown that the maximum peak ME voltage coefficient can be obtained for unclamped samples with 3-0 connectivity. Clamping leads to significant variation of ME voltage coefficients. Therefore for any comparison of theory and data one needs to take into account the degree of sample clamping. The results presented here are of practical importance for achieving bulk composite with desired ME parameters.

*Research at Novgorod State University was supported by grants from the Russian Ministry of Education (Grant No. E02-3.4-278) and from the Universities of Russia Foundation (Grant No. UNR 01.01.007). The National Science Foundation (Grant No. DMR-0302254) supported the efforts at Oakland University.*